\documentclass[twocolumn,aps,prd,showpacs,superscriptaddress,floatfix]{revtex4}

\usepackage{bm}
\usepackage{dcolumn}
\usepackage{graphicx}
\usepackage{epsf}
\usepackage{epsfig}

\usepackage{amsmath}
\usepackage{amsfonts}
\usepackage{amssymb}

\newcommand{\addrFR}{Theoretische Quantendynamik,
Physikalisches Institut der Universit\"{a}t Freiburg,\\
Hermann--Herder--Stra\ss{}e 3, 79104 Freiburg im Breisgau, Germany}

\newcommand{\addrNIST}{National Institute of Standards and Technology,
Gaithersburg, Maryland 20899--8401}

\begin{document}

\bibliographystyle{myprsty}

\title{Electron Self Energy for Higher Excited $S$ Levels}

\author{Ulrich~D.~Jentschura}
\affiliation{\addrFR}
\affiliation{\addrNIST}

\author{Peter~J.~Mohr}
\affiliation{\addrNIST}

\begin{abstract}
A nonperturbative numerical evaluation of the one-photon electron self
energy for the $3S$ and $4S$ states with 
charge numbers $Z=1$ to 5 is described. 
The numerical results are in agreement with known terms in 
the expansion of the self energy in
powers of $Z\alpha$.
\end{abstract}


\maketitle

In this brief report, we consider the one-loop self-energy shift 
which is the dominant radiative correction to the 
energy of hydrogenic bound states.
For high-precision spectroscopy, $S$ states are rather
important because they can be excited from the ground state
via Doppler-free two-photon spectroscopy.
We calculate the self-energy numerically to high accuracy
for $3S$ and $4S$ states (nuclear charge number
$Z=1,\dots,5$). We follow the approach previously 
outlined for $1S$~(Ref.~\cite{JeMoSo1999}) and 
$2S$ and $2P$ states~(Ref.~\cite{JeMoSo2001pra}).

The natural
unit system with $\hbar = c = m_{\rm e} = 1$ and $e^2 = 4\pi\alpha$ is
employed, as is customary in bound-state quantum
electrodynamics.
The (real part of the) energy shift $\Delta E_{\rm SE}$ due to the
electron self-energy radiative correction is usually written 
as~\cite{SaYe1990}
\begin{equation}
\label{ESEasF}
\Delta E_{\rm SE} = \frac{\alpha}{\pi} \, 
\frac{(Z \alpha)^4 \, m_{\rm e}}{n^3} \, 
F(nl_j,Z\alpha) \,,
\end{equation}
where $F$ is a dimensionless quantity. 
In writing the expression $F(nl_j,Z\alpha)$, we follow the usual
spectroscopic notation for an atomic state with 
principal quantum number
$n$, orbital angular momentum $l$ and 
total electron angular momentum $j$.  

\begin{table}[htb]
\begin{center}
\begin{minipage}{7.7cm}
\begin{center}
\caption{\label{tableF3S} Numerical results for
the scaled self-energy function
$F$ ($3S$ state) and the self-energy remainder function $G_{\rm SE}$,
in the regime of low nuclear charge numbers $Z$.}
\begin{tabular}{l@{\hspace*{0.5cm}}r@{.}l@{\hspace*{0.5cm}}r@{.}l}
\hline
\hline
$Z$ &
\multicolumn{2}{l}{\rule[-3mm]{0mm}{8mm}
  $F(3{S}_{1/2},Z\alpha)$} &
\multicolumn{2}{l}{\rule[-3mm]{0mm}{8mm}
  $G_{\rm SE}(3{S}_{1/2},Z\alpha)$}\\
\hline
\rule[-1mm]{0mm}{6mm}
$1$ & $10$ & $605~614~22(5)$ & $-31$ & $047~7(9)$ \\
\rule[-1mm]{0mm}{6mm}
$2$ &  $8$ & $817~615~14(5)$ & $-30$ & $512~6(2)$ \\
\rule[-1mm]{0mm}{6mm}
$3$ &  $7$ & $794~461~17(5)$ & $-30$ & $022~7(1)$ \\
\rule[-1mm]{0mm}{6mm}
$4$ &  $7$ & $083~612~42(5)$ & $-29$ & $564~53(6)$ \\
\rule[-3mm]{0mm}{8mm}
$5$ &  $6$ & $543~385~98(5)$ & $-29$ & $130~61(4)$ \\
\hline
\hline
\end{tabular}
\end{center}
\end{minipage}
\end{center}
\end{table}

The leading terms 
in the semi-analytic expansion of $F(n{S}_{1/2},Z\alpha)$
about $Z\alpha=0$ read
\begin{eqnarray}
\label{defFLOnS}
\lefteqn{F(n{S}_{1/2},Z\alpha) = 
A_{41}(n{S}_{1/2}) \, \ln(Z \alpha)^{-2}} \nonumber\\[2ex]
& & \;\; + A_{40}(n{S}_{1/2}) +
(Z \alpha) \, A_{50}(n{S}_{1/2}) \nonumber\\[2ex]
& & \;\; + \, 
(Z \alpha)^2 \, 
\left[A_{62}(n{S}_{1/2}) \, \ln^2(Z \alpha)^{-2} \right. 
\nonumber\\[2ex]
& & \;\;\;\; \left. + A_{61}(n{S}_{1/2}) \,\ln(Z \alpha)^{-2} + 
G_{\rm SE}(n{S}_{1/2},Z\alpha) \right]\,.
\end{eqnarray}
The $A$ coefficients have two indices, the first of which denotes the
power of $Z\alpha$ [including those powers implicitly contained in
Eq.~(\ref{ESEasF})], while the second index denotes the power of the
logarithm $\ln(Z \alpha)^{-2}$.  

We now list the analytic coefficients and the
Bethe logarithms relevant to the atomic states under
investigation~\cite{Be1947,Fe1948,Fe1949,FrWe1949,KrLa1949,Sc1949,FuMiTo1949,%,
Ba1951,KaKlSc1952,BaBeFe1953,FrYe1958,FrYe1960,La1960,La1961a,La1961b,%,
ErYe1965a,ErYe1965b,Er1971,Sa1981,Pa1993}, 
\begin{subequations}
\begin{eqnarray}
\label{coeffs1S12}
A_{41}(n{S}_{1/2}) & = & \frac{4}{3}\,, \\[1ex]
A_{40}(n{S}_{1/2}) & = & 
  \frac{10}{9} - \frac{4}{3} \, \ln k_0(n{S})\,, 
\\[1ex]
A_{50}(n{S}_{1/2}) & = & 
4\pi\,\left[\frac{139}{128} - \frac{1}{2}\,\ln 2\right]\,,
\\[1ex]
A_{62}(n{S}_{1/2}) & = & -1\,.
\end{eqnarray}
\end{subequations}
$A_{61}$-coefficients read
\begin{subequations}
\begin{eqnarray}
A_{61}(1{S}_{1/2}) & = & \frac{21}{20} + \frac{28}{3} \, \ln 2 \,,\\[1ex]
A_{61}(2{S}_{1/2}) & = & \frac{67}{30} + \frac{16}{3} \, \ln 2 \,,\\[1ex]
A_{61}(3{S}_{1/2}) & = & \frac{6163}{1620} + \frac{28}{3} \, \ln 2 - 4\,\ln 3\,,
\\[1ex]
A_{61}(4{S}_{1/2}) & = & \frac{4}{3} \, \ln 2 + \frac{391}{80}\,.
\end{eqnarray}
\end{subequations}
The Bethe logarithms $\ln k_0(n{S})$ are 
known~\cite{KlMa1973,BeBrSt1950,Ha1956,ScTi1959,Li1968,Hu1969},
and we here present a re-evaluation, 
\begin{subequations}
\begin{eqnarray}
\label{BetheLog1S}
\ln k_0(1{S}) &=& 2.984~128~555~765~497~611(1)\,,\\[1ex]
\label{BetheLog2S}
\ln k_0(2{S}) &=& 2.811~769~893~120~563~520(1)\,,\\[1ex]
\label{BetheLog3S}
\ln k_0(3{S}) &=& 2.767~663~612~491~821~190(1)\,,\\[1ex]
\label{BetheLog4S}
\ln k_0(4{S}) &=& 2.749~811~840~454~057~422(1)\,.
\end{eqnarray}
\end{subequations}
The results for $1S$, $2S$ and $3S$ are in agreement with values 
indicated for the ``logarithmic sum'' $\beta_1$ in~\cite[Tab.~III]{GoDr2000}.
Notice that the numerical difference of our above results
for $\ln k_0$ in comparison to the 
results indicated for $\beta_1$ in~\cite{GoDr2000} is entirely due to an 
additional contribution $\ln 2$ which is added to the 
Bethe logarithm of $S$ states according to the 
somewhat non-standard convention for $\beta_1$ used in~\cite{GoDr2000}.

\begin{table}[htb]
\begin{center}
\begin{minipage}{8.6cm}
\begin{center}
\caption{\label{tableF4S} Numerical results for
the scaled self-energy function
$F(4S_{1/2}, Z\alpha)$ and the self-energy remainder function $G_{\rm SE}$.}
\begin{tabular}{l@{\hspace*{0.5cm}}r@{.}l@{\hspace*{0.5cm}}r@{.}l}
\hline
\hline
$Z$ &
\multicolumn{2}{l}{\rule[-3mm]{0mm}{8mm}
  $F(4{S}_{1/2},Z\alpha)$} &
\multicolumn{2}{l}{\rule[-3mm]{0mm}{8mm}
  $G_{\rm SE}(4{S}_{1/2},Z\alpha)$}\\
\hline
\rule[-1mm]{0mm}{6mm}
$1$ & $10$ & $629~388~4(2)$ & $-30$ & $912(4)$\\
\rule[-1mm]{0mm}{6mm}
$2$ &  $8$ & $841~324~1(2)$ & $-30$ & $380~0(9)$ \\
\rule[-1mm]{0mm}{6mm}
$3$ &  $7$ & $818~078~5(2)$ & $-29$ & $892~4(4)$ \\
\rule[-1mm]{0mm}{6mm}
$4$ &  $7$ & $107~116~6(2)$ & $-29$ & $437~1(2)$ \\
\rule[-3mm]{0mm}{8mm}
$5$ &  $6$ & $566~758~8(2)$ & $-29$ & $006~0(2)$ \\
\hline
\hline
\end{tabular}
\end{center}
\end{minipage}
\end{center}
\end{table}

The evaluation of the coefficient 
\begin{equation}
A_{60}(nS_{1/2}) \equiv \lim_{Z\alpha \to 0} G_{\rm SE}(nS_{1/2},Z\alpha)
\end{equation}
has been
historically problematic~\cite{ErYe1965a,ErYe1965b,%,
Er1971,Sa1981,Pa1993}, and it has therefore been 
a considerable challenge to reliably estimate the 
self-energy remainder function $G_{\rm SE}$,
especially in the 
range of low nuclear charge number $Z$.
Our calculation of the nonperturbative (in 
$Z\alpha$) electron self-energy for the $3{S}_{1/2}$ state 
(see Table~\ref{tableF3S}) has 
a numerical uncertainty of 2~Hz in atomic
hydrogen. For the $4{S}_{1/2}$ state, the
numerical uncertainty is 3$\times Z^4$~Hz (see Table~\ref{tableF4S}).
The value of the fine-structure constant 
$\alpha$ employed in the calculation is
$\alpha^{-1} = 137.036$; this is close to the 
1998 and 2002 CODATA recommended values~\cite{MoTa2000,MoTaPriv2004}. 
The entries for the self-energy remainder 
function $G_{\rm SE}$ in Tables~\ref{tableF3S}
and~\ref{tableF4S} are in agreement with those used 
in the latest adjustment of the fundamental physical 
constants~\cite{MoTaPriv2004} (the $G_{\rm SE}$-values 
used in~\cite{MoTaPriv2004} are based on 
an extrapolation of numerical data previously
obtained~\cite{MoKi1992} for higher nuclear charge numbers). Our all-order
evaluation eliminates any uncertainty due to the unknown
higher-order analytic terms that contribute to 
the bound electron self-energy of $3S$ and $4S$ 
states [see Eq.~(\ref{defFLOnS})].
This improves our knowledge of the spectrum of
hydrogenlike atoms (e.g.~atomic hydrogen, He$^+$).\\[2ex]

{\bf Acknowledgments.}
U. D. J. thanks the National Institute of Standards and Technology for
kind hospitality during a number of extended research appointments. 
The authors acknowledge E.-O.~LeBigot for help in obtaining numerical
results for selected partial contributions to the electron self-energy,
for the hydrogenic energy levels discussed in this work.

\end{document}